\documentclass[a4paper,fleqn,usenatbib]{mnras}

\usepackage{newtxtext,newtxmath}
\usepackage[T1]{fontenc}
\usepackage{ae,aecompl}

\usepackage{graphicx}	% Including figure files
\usepackage{amsmath}	% Advanced maths commands
\usepackage{amssymb}	% Extra maths symbols
\usepackage{longtable}
\usepackage{pdflscape}

\title[A pilot survey of the binarity of MYSOs]{A pilot survey of the binarity of Massive Young Stellar Objects with $K$ band adaptive optics}

\author[R. Pomohaci et al.]{
Robert Pomohaci,$^{1}$\thanks{E-mail: rob.pomohaci@gmail.com}
Ren\'e D. Oudmaijer,$^{1}$ \thanks{E-mail: r.d.oudmaijer@leeds.ac.uk}
Simon P. Goodwin$^{2}$
\\
$^{1}$School of Physics and Astronomy, University of Leeds, Woodhouse Lane, Leeds LS2 9JT, UK\\
$^{2}$Department of Physics and Astronomy, University of Sheffield, Hicks Building, Hounsfield Road, Sheffield, S3 7RH, UK \\
}

\date{Accepted XXX. Received YYY; in original form ZZZ}

\pubyear{2018}

\begin{document}
\label{firstpage}
\pagerange{\pageref{firstpage}--\pageref{lastpage}}
\maketitle

% Abstract of the paper
\begin{abstract}
We present the first search for binary companions of Massive Young Stellar Objects (MYSOs) using AO-assisted $K$ band observations, with NaCo at the VLT. We have surveyed 32 MYSOs from the RMS catalogue, probing the widest companions, with a physical separation range of 400 - 46,000 au, within the predictions of models and observations for
multiplicity of MYSOs. Statistical methods are employed to discern whether these companions are physical rather than visual binaries.  We find 18 physical companions around 10 target objects, amounting to a multiplicity fraction of 31$\pm$8\% and a companion fraction of 53$\pm$9\%. For similar separation and mass ratio ranges, MYSOs seem to have more companions than T Tauri or O stars, respectively. This suggests that multiplicity increases with mass and decreases with evolutionary stage.  We compute very rough estimates for the mass ratios from the {\it K} band magnitudes, and these appear to be  generally larger than 0.5. This is inconsistent with randomly sampling the IMF, as predicted by the binary capture formation theory. Finally, we find that MYSOs with binaries do not show any different characteristics to the average MYSO in terms of
luminosity, distance, outflow or disc presence.

\end{abstract}

% Select between one and six entries from the list of approved keywords.
% Don't make up new ones.

%%%%%%%%%%%%%%%%%%%%%%%%%%%%%%%%%%%%%%%%%%%%%%%%%%

%%%%%%%%%%%%%%%%% BODY OF PAPER %%%%%%%%%%%%%%%%%%

%\citet{moe2017}

\section{Introduction}
\label{sec:naco:intro}
Studying the star formation process is of crucial importance to many branches of astrophysics, from stellar evolution to cosmology. Low-mass star formation is thought to result from the monolithic collapse of a cloud followed by accretion through the circumstellar disc, and is in general reasonably well understood as per the description of \citet{krumholz14}. However, the way in which massive stars form poses a number of challenging problems. As shown by the simulations of \citet{kahn74}, radiation pressure halts spherical accretion for objects more massive than 40 M$_{\odot}$. Given that observations have found stars of 100 M$_{\odot}$ and above (\citealt{Crowther16}), there must be a way for radiation to escape without halting the accretion process.\\
There are two main theoretical approaches to explaining massive star formation: monolithic collapse of turbulent cores (\citealt{mckee03}) and competitive accretion (\citealt{Bonnell+Bate2006}). Monolithic collapse suggests that high-mass stars form in a similar fashion to low mass stars, in clumps supported against collapse by turbulence, with different mechanisms being used to eliminate radiation pressure such as ionised jets and winds. Competitive accretion takes the different view that massive stars form only in the centers of clusters, where they can take advantage of the stronger gravitational potential.\\
The binarity of the Massive Young Stellar Objects (MYSOs) could be a  fundamental aspect of massive star formation, and is predicted by both the monolithic collapse and competitive accretion scenarios. This binarity could be due to disc fragmentation (\citealt{krumholz09}), capture (\citealt{moeckel07}),  or a result of the dense environment in competitive accretion (\citealt{Bonnell+Bate2006}).  Recently, \citet{lund2018} proposed that close high mass binaries can result from accreting low mass stars in wide binaries. Whereas the monolithic collapse scenario predicts binary separations of order 1000s of au (\citealt{krumholz12}, but see \citealt{rosen16}), the captured binaries in competitive accretion quickly become close at separations of less than 10 au (\citealt{bonnell05}).\\
The binary fractions and properties of MS (Main-Sequence) stars will not necessarily reflect the primordial binary properties as the latter will have been affected by secular evolution such as dynamical processes (see the review by \citealt{kratter11}).  An important question to address is how these binaries were formed and evolved. It is interesting to note that although as mentioned above, various theories are capable of forming binaries, they have not been put to the test using observations and have not even been informed by targeted observations.\\
Several surveys of Main-Sequence stars have been undertaken, and they generally find that multiplicity increases with stellar mass. Field A-stars were surveyed by \citet{derosa14}, who were able to probe a very large separation range of 30 and 45000\,au to find an overall multiplicity fraction of 43.6$\pm$3.4\%. Whereas for solar-mass stars the fraction is less that half (46\% according to \citealt{raghavan10}), \citet{sana12} found multiplicity frequencies of over 70\% for O stars.  \citet{moe2017} extended this result in a very comprehensive study. They found that the mass ratio of binary systems with early type primaries appear to favour values around 0.5 for the closest companions, but that the mass ratios for the companions at large separations (200-5000 au) were consistent with random sampling from the Initial Mass Function (IMF). \\
In addition, in many cases the binary pairs are close enough for interactions to occur at some point in the evolution of the star, be it through envelope stripping, accretion and spin up, common envelope evolution or merging. This has significant consequences for stellar evolution models, as most of these are based on single stars. Cluster evolution is also shaped by stellar interactions (\citealt{parker13}). As a result the evolution and fate of a massive star is also governed by its binary properties, rather than only by its initial mass as in the single star evolutionary scenarios (see eg. \citealt{smith15}). \\
Binary surveys of low-mass Class II/III YSOs in the Taurus cluster have found a very high companion frequency (65-80\%, \citealt{kraus11}). Mass ratio distributions seem to be fairly flat. By contrast, dense clusters like the ONC show much lower companion frequencies similar to the field (\citealt{reipurth08}, but see \citealt{duchene18}). \\
The closest observations to massive pre-main-sequence stars are the spectro-astrometry studies of \citet{baines06} and \citet{wheelwright10} on the intermediate mass Herbig Ae/Be stars. They found high multiplicity frequencies (over 70\%) and high mass ratios, close to equal mass companions. This is inconsistent with random sampling from the IMF, which is a prediction of the capture theory. They also found higher mass Herbig Be stars to have larger multiplicity fractions than Herbig Ae stars (when uncertain detections were not included), same as the trend seen for MS stars. Finally, binary orbits and disc planes were found to be coplanar at a 2.2-sigma level, based on comparisons with simulated distributions, and highly inconsistent with random orientations. This gives tentative support to the disk fragmentation formation scenario. \\
For the embedded MYSOs,  many of the studies have been single-object and often concerned a serendipitous discovery. Using AMBER VLT-I in the $H$ and $K$-band, \citet{kraus12v921} discovered a close, 29 au, companion to the Herbig Be star V921 Sco, \citet{kraus06} used K-band speckle interferometry and discovered a binary companion at a distance of 195 mas ($\sim$700 au) from the high mass protostar NGC7538 IRS2. \citet{apai07} studied a sample of 16 embedded O stars, searching for radial velocity differences in order to detect binary companions. They found that two of their targets showed velocity differences of 90 km/s between the two different epochs, interpreted as being caused by close binary companions. \citet{beuther17} studied the massive protostar/UCH{\sc ii} region NGC 7538 IRS1 using JVLA data, discovered a binary source at 430 au, and found (misaligned) disks surrounding both objects. \citet{kraus17} discovered a companion at 58 mas (170 au) from the 20M$_{\odot}$ protostar IRAS 17216-3801. They determined the masses of the two components to be 20 (primary) and 18 M$_{\odot}$ (secondary) using the K-band flux ratios, and also found misaligned disks. Even closer binary companions were found in the VLT-I data of Koumpia et al. (A\&A submitted) of two massive young stars.  Recently, \citet{sana17} observed 17 MS and pre-MS stars in M17, and found a low radial velocity dispersion ($\sigma = 5.6 \pm 0.2 $ km/s). They interpret this as support for the idea that binaries form at large separations, and then come close due to tidal interaction throughout their lifetimes.\\
\begin{figure}
	\includegraphics[scale=0.1]{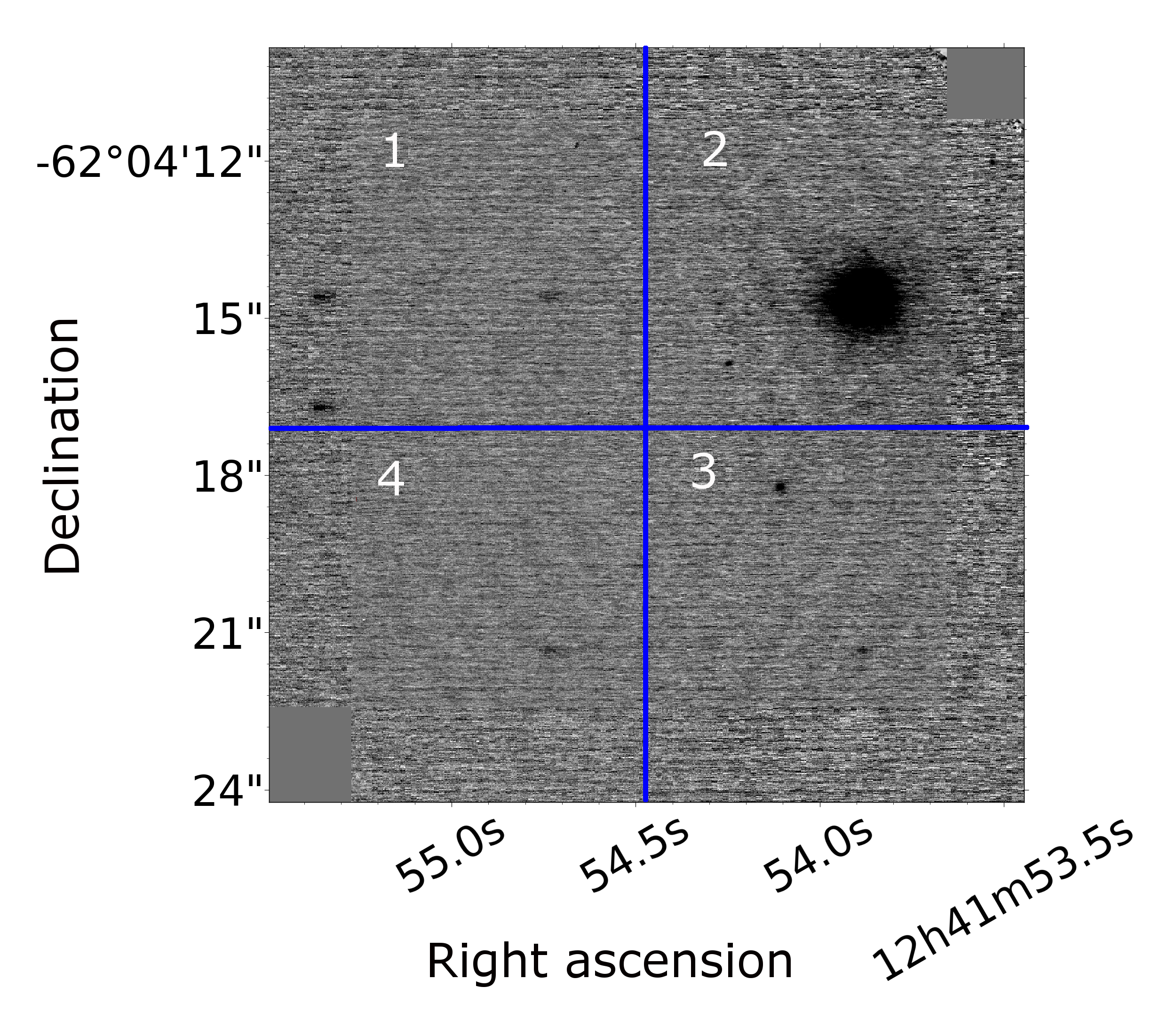}
    \caption{Reduced image of G301.8147, detailing the set up of the observations. The primary MYSO target is centered in quadrant 2, in the upper right part of the image, due to a technical fault that affected quadrant 4. The area of quadrant 2 is where we focus our search for companions.}
    \label{fig:quadrants}
\end{figure}
However, there has been no targeted survey to date of the multiplicity of massive pre-main sequence stars.  This paper describes a pilot adaptive optics survey aimed at searching for MYSO binary companions. The targets were selected from the RMS survey, a multi-wavelength galaxy-wide search for MYSOs (see \citet{lumsden13} for a description of the survey and its results). The observations presented in this paper are the first step of a larger project to determine the binary statistics of  massive young pre-Main-Sequence objects. Section \ref{sec:naco:obs} describes the sample selection and the data reduction process. Section \ref{sec:naco:results} presents the initial
results - an analysis of the completeness limits of the survey, the methods employed to eliminate visual binaries and preliminary binary statistics. A discussion on the multiplicity fraction, mass ratios, period distributions, disc-binary orbit alignment and whether the binaries are different to the Galactic population of MYSOs as seen by the RMS survey is presented in Section \ref{sec:naco:discussion}. The conclusions of this work are presented in Section \ref{sec:naco:concl}.

\begin{landscape}
\setlength\LTcapwidth{\linewidth}
\begin{table}
\caption[NaCo source list]{Source list, including observation conditions and quality. Right Ascension and Declination correspond to the RMS survey position of the MYSO. 1 - Magnitudes from 2MASS; 2 - distances from \citet{urquhart11a}, they carry an uncertainty of order 0.5-1 kpc; 3 - luminosities calculated by \citet{mottram11}. The uncertainties on bolometric fluxes are of the order 10-20\%. Stellar masses are determined from bolometric luminosities  by logarithmic interpolation of the pre-MS relations of \citet{Davies2011}.  Combined with distance uncertainties and those in bolometric luminosity, the masses can have uncertainties of order 50\%; 'Mult?' indicates whether the targets have any detected companions (a list of these is presented in Table \ref{tab:naco:comps}); SR is the Strehl Ratio of the given observation; the limiting magnitudes are determined as detailed in Section \ref{sec:naco:completeness}} \label{tab:naco:obslog}
 \begin{tabular}{llllllllllrrllrl}
 \hline
Date       & Object name        & Mult? & Texp & RA       & Dec         & $J$  & $H$  & $K$  & D     & Lbol          & Mass          & Airm. & Seeing & SR     & Lim  \\
           &                    &       & (s)  & (J2000)  & (J2000)     &      &      &      & (kpc) & (L$_{\odot}$) & (M$_{\odot}$) & (")   & mag    &        & mag  \\
           \hline
20.12.2015 & G212.0641-00.7395  & N     & 160  & 06:47:13 & +00:26:06.5 & 14.3 & 12   & 10   & 4.7   & 16000         & 14.5          & 1.33  & 0.75   & 5.3\%  & 13.5 \\
20.12.2015 & G221.9605-01.9926  & Y     & 160  & 07:00:51 & -08:56:30.1 & 14.2 & 11.4 & 9.2  & 3.2   & 5500          & 10            & 1.22  & 0.67   & 14.7\% & 13.4 \\
25.12.2015 & G231.7986-01.9682  & N     & 120  & 07:19:36 & -17:39:18.0 & 9.2  & 7.8  & 6.4  & 3.2   & 5600          & 10.1          & 1.4   & 1.6    & 25.1\% & 14.5 \\
25.12.2015 & G232.0766-02.2767  & Y     & 160  & 07:19:00 & -18:02:41.6 & 13.4 & 12   & 10.4 & 3     & 5000          & 9.7           & 1.01  & 1.1    & 4.7\%  & 13.3 \\
27.12.2015 & G232.6207+00.9959  & N     & 160  & 07:32:10 & -16:58:13.4 & 13.1 & 12.4 & 8.3  & 1.7   & 11000         & 12.5          & 1.21  & 0.75   & 21.4\% & 14.8 \\
25.12.2015 & G233.8306-00.1803  & N     & 120  & 07:30:17 & -18:35:49.1 & 10.9 & 8    & 6.1  & 3.3   & 13000         & 13            & 1.02  & 1.1    & 25.0\% & 13.9 \\
22.12.2015 & G268.3957-00.4842  & Y     & 160  & 09:03:25 & -47:28:27.5 & 15.7 & 11.8 & 8.3  & 0.7   & 3000          & 7.8           & 1.22  & 0.71   & 21.2\% & 15.3 \\
27.12.2015 & G282.2988-00.7769  & Y     & 120  & 10:10:00 & -57:02:07.3 & 10.2 & 8.4  & 7    & 3.7   & 4000          & 8.6           & 1.19  & 1.5    & 50.0\% & 14.5 \\
04.01.2016 & G287.3716+00.6444  & Y     & 160  & 10:48:05 & -58:27:01.5 & 10.4 & 8.9  & 7.5  & 4.5   & 18000         & 14.9          & 1.21  & 0.85   & 40.0\% & 14.3 \\
04.01.2016 & G290.3745+01.6615  & Y     & 160  & 11:12:18 & -58:46:20.8 & 12.2 & 10   & 8.6  & 2.9   & 15000         & 14            & 1.21  & 1.3    & 20.4\% & 13.4 \\
10.02.2016 & G293.5607-00.6703  & N     & 160  & 11:30:07 & -62:03:12.8 & 14.9 & 12.2 & 9.5  & 3.4   & 4000          & 8.6           & 1.27  & 1.2    & 3.0\%  & 12.7 \\
10.02.2016 & G300.1615-00.0877  & N     & 160  & 12:27:09 & -62:49:44.2 & 15.7 & 12.1 & 9.3  & 4.2   & 5600          & 10.1          & 1.39  & 1.3    & 13.7\% & 14.9 \\
10.02.2016 & G300.3412-00.2190  & N     & 160  & 12:28:36 & -62:58:35.4 & 13.3 & 10.7 & 8.7  & 4.2   & 6000          & 10.8          & 1.46  & 1.4    & 19.6\% & 14.9 \\
10.02.2016 & G301.1726+01.0034  & N     & 160  & 12:36:32 & -61:49:02.8 & 12.7 & 10.2 & 7.9  & 4.3   & 21000         & 15.8          & 1.27  & 1.55   & 23.1\% & 14.3 \\
10.02.2016 & G301.8147+00.7808A & Y     & 160  & 12:41:54 & -62:04:14.6 & 12   & 9.3  & 6.8  & 4.4   & 22000         & 16.1          & 1.29  & 0.95   & 24.9\% & 14.2 \\
10.02.2016 & G305.2017+00.2072A & N     & 160  & 13:11:10 & -62:34:38.6 & 14.2 & 11.7 & 9.4  & 4     & 30000         & 18            & 1.27  & 0.9    & 12.4\% & 13.8 \\
10.02.2016 & G305.3676+00.2095  & Y     & 160  & 13:12:36 & -62:33:32.3 & 14.8 & 13.1 & 10.4 & 4     & 16000         & 14.5          & 1.27  & 1.2    & 4.8\%  & 14.4 \\
10.02.2016 & G305.5610+00.0124  & N     & 160  & 13:14:26 & -62:44:30.4 & 15.7 & 12.6 & 9.7  & 4     & 12000         & 12.8          & 1.28  & 0.85   & 10.3\% & 12.3 \\
10.02.2016 & G305.6327+01.6467  & N     & 120  & 13:13:48 & -61:06:28.8 & 8.6  & 7.5  & 7.2  & 4.9   & 16000         & 14.5          & 1.3   & 1.1    & 24.6\% & 14.4 \\
12.02.2016 & G309.9796+00.5496  & N     & 160  & 13:51:03 & -61:30:14.1 & 15.9 & 12.4 & 9.7  & 3.5   & 7600          & 11.2          & 1.28  & 0.9    & 10.1\% & 13.5 \\
10.02.2016 & G310.0135+00.3892  & Y     & 191  & 13:51:38 & -61:39:07.5 & 11.8 & 7.6  & 4.9  & 3.2   & 67000         & 23.6          & 1.45  & 1.1    & 24.9\% & 15.1 \\
12.02.2016 & G311.4402+00.4243  & N     & 160  & 14:03:07 & -61:15:27.9 & 14.3 & 10.3 & 7.8  & 3.6   & 7100          & 10.9          & 1.25  & 1.05   & 23.5\% & 12.7 \\
10.02.2016 & G320.1542+00.7976  & N     & 160  & 15:05:17 & -57:31:40.0 & 11.2 & 10.3 & 9.8  & 2.5   & 5400          & 9.5           & 1.38  & 1.7    & 9.2\%  & 13.8 \\
10.02.2016 & G326.4755+00.6947  & Y     & 160  & 15:43:19 & -54:07:35.4 & 15.4 & 12.4 & 9.3  & 1.8   & 4100          & 8.6           & 1.49  & 1.5    & 13.7\% & 13.9 \\
10.02.2016 & G327.9455-00.1149  & N     & 160  & 15:54:35 & -53:50:42.1 & 15.8 & 12.5 & 10   & 3.1   & 4300          & 9.3           & 1.38  & 1.4    & 5.3\%  & 13.4 \\
20.03.2016 & G331.3576+01.0626  & N     & 160  & 16:06:26 & -50:43:22.0 & 12.6 & 11.1 & 9.6  & 4.5   & 18000         & 15            & 0.7   & 1.33   & 11.8\% & 15.1 \\
10.03.2016 & G332.0939-00.4206  & N     & 120  & 16:16:16 & -51:18:25.2 & 15.3 & 9.6  & 5.9  & 3.6   & 93000         & 28            & 1.28  & 1.4    & 25.1\% & 13.5 \\
17.03.2016 & G332.9868-00.4871  & N     & 160  & 16:20:38 & -50:43:49.6 & 17.6 & 13.7 & 9.3  & 3.6   & 18000         & 15            & 1.24  & 1.25   & 12.8\% & 13   \\
20.03.2016 & G334.7302+00.0052  & N     & 160  & 16:26:05 & -49:08:41.8 & 15.5 & 13   & 9.6  & 2.5   & 3800          & 8.5           & 1.48  & 0.9    & 2.6\%  & 14.1 \\
17.03.2016 & G336.4917-01.4741B & N     & 160  & 16:40:01 & -48:51:52.4 & 11.7 & 10.3 & 8.8  & 2     & 12000         & 14            & 1.24  & 1.2    & 19.1\% & 14.1 \\
17.03.2016 & G339.6816-01.2058  & N     & 160  & 16:51:06 & -46:15:52.4 & 13.1 & 10.4 & 8.5  & 2.4   & 6500          & 11            & 1.21  & 1.1    & 20.5\% & 13.6 \\
17.03.2016 & G344.8889+01.4349  & N     & 120  & 16:57:52 & -40:33:26.7 & 14.1 & 10.2 & 7.4  & 2.4   & 7000          & 11.3          & 1.16  & 1.1    & 24.4\% & 14.5 \\
\hline
 
\end{tabular}
\end{table}

\end{landscape}
\noindent
\section{Observations}
\label{sec:naco:obs}

We selected 32 MYSO targets from the RMS survey. This study was a multi-wavelength effort to detect all MYSOs in the Galaxy. RMS employed multi-wavelength observations in order to distinguish candidate MYSOs from similar-looking sources such as evolved stars, compact HII regions, planetary nebulae, etc. The final survey found 800 MYSOs and H{\sc ii} regions, and was determined to be complete to 18 kpc for massive young embedded sources brighter than 2$\times$10$^{4}$L$_{\odot}$ (except the inner 20$^{\circ}$ of
Galactic longitude which were omitted due to source confusion). 
The final results from the survey are compiled in \citet{lumsden13}. The target selection for this NACO survey was based on bolometric luminosity and distance cuts, as well as declinations accessible from the ESO-Very Large Telescope (VLT). All but one of the objects are classified as MYSOs in the final RMS catalogue. G331.3576 was initially classified as an HII region, due to its proximity to the well-known HII region G331.3546. However, subsequent near-infrared spectra (Lumsden et al., in prep.) have shown that G331.3576 is a featureless central star exciting a compact HII region. The observed objects have L$>$3500 L$_{\odot}$ (corresponding to M$>$9 M$_{\odot}$ according to the
mass-luminosity relation from \citealt{Davies2011}), d$<$5 kpc (for the highest completeness of the RMS survey), $\delta <$10$\rm ^{o}$ (to be easily observed from the VLT), and $K<$10.5 mag (so that the targets can be their own guide stars for the adaptive optics correction).  \\ 
The $K$ band was chosen as this is the shortest wavelength (allowing the highest spatial resolution) at which all of the heavily extincted MYSOs are bright enough to be still visible with short on-target times. \\ 
The survey probes the separation range from the minimum achieved FWHM of the images of $\sim$120 mas out to the full field of view of 14$\times$14 arcseconds. For the average distance of MYSOs in this sample of 3.3 kpc, this means binaries with separations from 400 - 46,000 au can be resolved. For comparison, the massive binary separation range from simulations of \citet{krumholz09} and \citet{krumholz12} was 1590 au, while the sparse observations in the literature indicate separations of 400-700 au. The average $K$ band magnitude of the targets is 9 mag. For Main Sequence stars, for $\Delta K$ = 6, companions up to 15 times less massive can in principle be recovered (as per Fig. 4 of \citealt{oudmaijer10}). However, this can be affected by differential dust excess emission and extinction between binary system components, so in practice the limiting mass ratio is likely lower than 1:15.\\
The observations were carried out in service mode between 20th Dec 2015 - 17th March 2016, with the AO system NaCo (\citealt{lenzen03}, \citealt{rousset03}) on the VLT at ESO. The diffraction limit of this instrument is 0.0057" or 57 mas (under ideal weather conditions and with perfect AO correction). NaCo has a 1024$\times$1024 pixel InSb camera, with a pixel size of 27 $\mu$m. In order to obtain the highest spatial resolution, the S13 camera mode was used, with a plate scale of 13.27 mas/pixel, resulting in a field of view of 13.6$\times$13.6 arcsec$^{2}$.\\
\begin{figure}
	\includegraphics[width=\columnwidth]{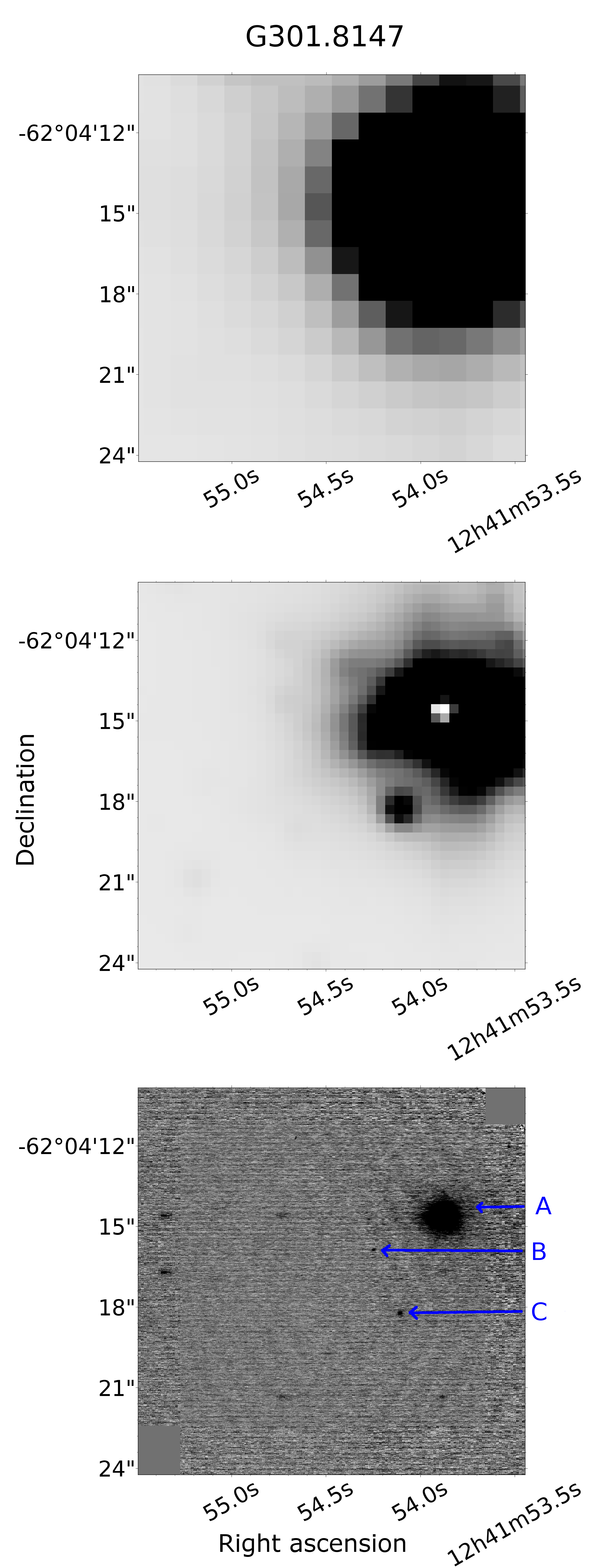}
    \caption{ Top - 2MASS image of MYSO G301.8147A. Middle - the same image from the VVV survey; Bottom - corresponding reduced image from NaCo. The three sources in the NaCo image are indicated by blue arrows. Other black lines visible are image artefacts. Note the new secondary source previously undetected in VVV or 2MASS.}
    \label{fig:exampledata}
\end{figure}
\newpage
\setlength\LTcapwidth{\linewidth}
\begin{landscape}
\begin{table}
\caption[Detected companions NACO]{All detected companions, along with their separations, position angles and $K$ band magnitudes determined relative to the primaries and 2MASS and VVV $K$ band magnitudes when previously detected.  Errors on separation are based on the image quality (as a result of the success of the AO correction), and they were then subsequently used to determine the uncertainties on PAs.'$\delta$m' is the difference in magnitude between the secondary and the primary. The 2MASS magnitude flag is a three-letter code (for the $J$, $H$ and $K$ bands ) which indicates the photometric quality of the observation. 'A' corresponds to detections with valid measurements at a 10$\sigma$ level, 'B' at 7$\sigma$, 'C' to 5$\sigma$ and 'D' to a measurement with no $\sigma$ requirement. 'U' is an upper limit on the magnitude. The three flags indicate the quality of the $J$, $H$ and $K$ band 2MASS observations in this order.} \label{tab:naco:comps} 
\begin{tabular}{lllllllllll}
    \hline
Object name                 & Sep   & $\Delta$Sep & PA       & $\Delta$PA & $K$   & $\Delta K$ &$\delta$m   & 2MASS     & 2MASS & VVV       \\
                           & (")   & (")         & ($\rm ^{o}$) & ($\rm ^{o}$)   & (mag) & (mag)      & (mag) & $K$ (mag) &      flag      & $K$ (mag) \\
                      \hline

G221.9605-01.9926B     & 0.60  & 0.26        & 75       & 25         & 10.8  & 0.1        & 1.6   &           &            &           \\
G221.9605-01.9926C     & 0.70  & 0.26        & 254      & 21        & 10.6  & 0.1        & 1.4   &           &            &           \\
G221.9605-01.9926D    & 1.10  & 0.26        & 60       & 13         & 10.2  & 0.1        & 1.0   &           &            &           \\
G221.9605-01.9926E     & 1.20  & 0.26        & 254      & 12         & 11.3  & 0.1        & 2.1   &           &            &           \\
G232.0766-02.2767B  & 3.13  & 0.24        & 178      & 4          & 9.6   & 0.2        & 1.7   & 9.8       & AEU        &           \\
G232.0766-02.2767C     & 5.58  & 0.24        & 153      & 2          & 9.3   & 0.2        & 1.4   & 11.3      & UUE        &           \\
G232.0766-02.2767D     & 4.30  & 0.24        & 206      & 3          & 7.5   & 0.1        & 2.5   & 10.8      & UUE        &           \\
G232.6207+00.9959B     & 6.43  & 0.17        & 77       & 1          & 13.3  & 0.1        & 5.0   & 10.1      & UDU        &           \\
G232.6207+00.9959C    & 6.66  & 0.17        & 125      & 4          & 10.9  & 0.1        & 2.6   & 11.2      & UUE        &           \\
G268.3957-00.4842B   & 1.92  & 0.12        & 350      & 1          & 11.7  & 0.2        & 3.4   &           &            &           \\
G268.3957-00.4842C    & 4.75  & 0.12        & 3        & 1          & 14.3  & 0.1        & 6.0   &           &            &           \\
G268.3957-00.4842D    & 8.70  & 0.12        & 94       & 1          & 14.4  & 0.1        & 6.1   &           &            &           \\
G268.3957-00.4842E    & 8.91  & 0.12        & 95       & 1          & 12.5  & 0.1        & 4.2   & 12.3      & AAA        &           \\
G282.2988-00.7769B     & 1.57  & 0.12        & 306      & 4          & 14.0  & 0.1        & 7.0   &           &            &           \\
G282.2988-00.7769C   & 2.71  & 0.12        & 200      & 3          & 14.5  & 0.1        & 7.5   &           &            &           \\
G287.3716+00.6444B    & 1.27  & 0.11        & 121      & 5          & 12.3  & 0.1        & 4.8   &           &            &           \\
G287.3716+00.6444C    & 1.48  & 0.11        & 112      & 4          & 13.5  & 0.2        & 6.0   &           &            &           \\
G287.3716+00.6444D     & 1.82  & 0.11        & 257      & 3          & 14.7  & 0.2        & 7.2   &           &            &           \\
G287.3716+00.6444E    & 1.90  & 0.11        & 28       & 3          & 12.7  & 0.1        & 5.2   &           &            &           \\
G290.3745+01.6615B    & 0.70  & 0.12        & 153      & 10         & 11.1  & 0.1        & 2.4   &           &            &           \\
G290.3745+01.6615C   & 1.88  & 0.12        & 156      & 4          & 11.9  & 0.1        & 3.2   &           &            &           \\
G290.3745+01.6615D    & 4.06  & 0.12        & 244      & 2          & 13.2  & 0.1        & 4.5   &           &            &           \\
G293.5607-00.6703B    & 5.69  & 0.29        & 339      & 3          & 11.1  & 0.2        & 1.5   & 11.1      & AEE        &           \\
G293.5607-00.6703C    & 4.60  & 0.29        & 90       & 4         & 13.7  & 0.1        & 4.1   &           &            &           \\
G300.3412-00.2190B     & 7.41  & 0.18        & 163      & 1          & 12.3  & 0.1        & 3.7   & 11.7      & AAA        & 11.9      \\
G301.8147+00.7808A\_B & 2.88  & 0.14        & 115      & 3          & 13.9  & 0.2        & 7.1   &           &            &           \\
G301.8147+00.7808A\_C  & 3.91  & 0.14        & 155      & 2          & 12.5  & 0.1        & 5.7   &           &            & 11.3      \\
G305.3676+00.2095B    & 0.88  & 0.15        & 288      & 10         & 12.7  & 0.1        & 2.3   &           &            &           \\
G310.0135+00.3892B    & 2.56  & 0.15        & 41       & 3          & 11.2  & 0.2        & 6.3   &           &            &           \\
G320.1542+00.7976B   & 2.95  & 0.21        & 162      & 4          & 10.4  & 0.1        & 0.5   & 9.8       & UUA        &           \\
G326.4755+00.6947B    & 2.19  & 0.23        & 179      & 6          & 13.5  & 0.1        & 4.2   &           &            &           \\
G327.9455-00.1149B   & 7.35  & 0.21        & 143      & 2          & 14.5  & 0.2        & 4.5   & 13.7      & CAA        & 13.9      \\
G331.3576+01.0626B  & 3.80  & 0.17        & 54       & 2         & 13.4  & 0.1        & 3.8   &           &            & 12.8      \\
G331.3576+01.0626C     & 7.80  & 0.17        & 193      & 1          & 12.4  & 0.1        & 2.8   & 12.0      & AAA        & 12.5      \\
G331.3576+01.0626D   & 11.25 & 0.17        & 133      & 1          & 12.9  & 0.1        & 3.3   & 12.0      & AAA        & 12.6      \\
G331.3576+01.0626E   & 11.41 & 0.17        & 109      & 1          & 12.5  & 0.1        & 2.9   & 12.2      & AAA        & 12.1      \\
G332.9868-00.4871B    & 9.14  & 0.23        & 201      & 1          & 12.1  & 0.1        & 2.8   & 11.4      & AAA        & 11.5      \\
G334.7302+00.0052B    & 5.79  & 0.24        & 157      & 2          & 12.9  & 0.1        & 3.3   & 12.0      & UAA        & 12.3      \\
G336.4917-01.4741B\_B & 4.88  & 0.21        & 98       & 2          & 12.4  & 0.1        & 3.6   & 11.7      & CAB        & 12.2      \\
G339.6816-01.2058B   & 8.59  & 0.15        & 85       & 1          & 12.7  & 0.1        & 4.2   & 13.2      & UUB        & 13.1     \\
\hline
\end{tabular}
\end{table}
\end{landscape}

The $Ks$ broadband filter was chosen in order to probe down to the lowest possible limiting magnitude in the shortest exposure times. This filter is centered on 2.18 $\mu$m and has a width of 0.35 $\mu$m. Due to a technical fault during the observing period, 1 in 8 columns in the lower left quadrant of the NaCo detector had no signal. As such, the targets were centered in the upper right quadrant, achieving full coverage only for distances up to $\sim$3" away from the main source. Figure \ref{fig:quadrants} details the set-up of the observations, with the MYSO centered in quadrant 2, which is the area that we focus our companion search in. For each of the targets, two object frames were taken, as well as two sky frames at 6" away from the main target. The exposure times were 80s per frame for targets of magnitude $K>$7.5, and shorter for brighter targets in order to avoid saturation. The seeing (in the $V$ band) was 1.7" or smaller, with an average of 1.1". The average airmass was 1.3. Information about the observations is presented in Table \ref{tab:naco:obslog}. \\

\noindent
The data were reduced with the ESO pipelines interfaced through the GASGANO software. Bad pixel maps, dark, twilight and lamp flat-field frames were provided for each observation. The dark current subtraction and flat field division were performed first, followed by bad pixel correction and sky subtraction. As it turned out, the fault in the 4th quadrant did not hamper the detection of sources in this part of the image. We cross-checked the GASGANO pipeline reduction with manual PyRAF routines, and the results from these two methods were consistent. \\
The average FWHM of the point sources is 0.12", and the average Strehl ratio 18\%. An example of the data is shown in Fig \ref{fig:exampledata}.\\

\subsection{Sensitivity - limiting magnitude and separation}
\label{sec:naco:completeness}
The first step in analysing the sensitivity of the data was to calculate the limiting magnitude of the observations. This was done by placing an artificial 2D Gaussian source of the same FWHM as the main target in an empty part of the image. For this we used the Gaussian2DKernel function in the AstroPy.convolution package (\citealt{astropy}).\\
The minimum flux at which the artificial source was detected at the 3$\sigma$ level above background noise by SExtractor was taken as the limiting flux. The limiting magnitude was then calculated by comparing the limiting flux to the primary flux with 2MASS photometry.\\
The average limiting magnitude determined as explained in the previous paragraph is 14 mag, close to the 2MASS survey limits, but not quite as deep as VVV. The data probe 5.5 mag fainter than the main source on average, but at higher spatial resolution than both VVV and 2MASS. The minimum separation at which secondary sources can be detected is determined by the quality of the AO correction. In order to determine the detection limit, we placed sources of random brightness at random distances and position angles around the main target. Three different MYSO observations were used for this, G233.8306 (FWHM of 0.08" and $\delta$m=7.5), G331.3576 (FWHM=0.11", $\delta$m=5) and G305.3676 (FWHM=0.18", $\delta$m=3). They were chosen as they are representative of the range of FWHM and magnitude differences of this survey. The limiting magnitude is constant at distances larger than 1". At shorter distances, the limiting magnitude becomes brighter (up to K$_{lim} \approx$=7 mag) due to the proximity of the K-band bright MYSO. \\
The minimum distance at which objects can be detected depends on the FWHM of the observation, and can be empirically quantified by d$_{\lim}$=1.5$\times$FWHM.\\
We searched the 2MASS and VVV point source catalogues for other sources that were not detected in the NaCo images as a consistency check. Most of the missed secondary objects are either fainter than the limiting magnitude, close to the edges of the NaCo field, or untrustworthy detections as indicated by survey quality flags. 13 sources did not fit any of the above criteria, and as such were visually inspected. All of these were either part of extended emission or close enough to the limiting magnitude to explain their non-detection in the NACO images.\\
In conclusion, the limiting magnitude ranges between $K$=12-15 mag for distances up to 1". At closer separations the limiting magnitude decreases, until d$\approx$1.5$\times$FWHM, which is the minimum distance at which a target object can be observed. The sample can be considered complete for $\Delta K$=5 mag at 1-3", our main area of interest and $\Delta K$=3 mag at 0.3-1".

\section{Results}
\label{sec:naco:results}
\subsection{Object detection}

We used the SExtractor software (\citealt{sextractor}) in combination with the GAIA-STARLINK package to identify sources in reduced frames. Sources at 3$\sigma$ or more above the background noise were taken as detections. SExtractor provides an estimate of the magnitude of an object (MAG\_AUTO), based on the first moment method of \citet{kron80}. 
\\
The magnitudes of the MYSOs in the RMS database were taken from 2MASS (\citealt{2mass}). The magnitudes of the newly detected sources were computed relative to the brightness of the main sources. Although the higher sensitivity VISTA Variables in the Via Lactea (VVV, \citealt{vista}) survey data was also available for two-thirds of the sources, we opted for 2MASS photometry as VVV saturates for magnitudes brighter than 11 in the $K$ band. \\
All of the targets are brighter than 10.5 magnitudes, so 2MASS primary magnitudes are more reliable. However, comparisons with VVV can still be drawn for secondary fainter sources present in this catalogue. Where available, secondary magnitudes as derived relative to the primaries generally agree with 2MASS catalogue magnitudes. Of note is the discrepancy of more than three magnitudes between G232.0766D, measured to be 7.5 magnitudes in the NaCo image, but catalogued at 10.8 magnitudes in 2MASS. This may be explained by variability in either this secondary source, or the primary MYSO. \\
The astrometry is calibrated relative to the RMS catalogue coordinates of the main source. The initial astrometry solution resulting from the GASGANO data reduction required small shifts to reconcile with the RMS coordinates, to correct for the pointing of the telescope. \\
We found a total of 40 secondary sources in the NaCo images, 21 of which are new detections, i.e. not present in either 2MASS or VVV survey data. The complete list of secondary sources, along with their parameters are presented in Table \ref{tab:naco:comps}. The new discoveries are closer to the primary targets (average separation 2.0") than the secondary objects already seen in survey data (average 6.4"). New detections are also fainter relative to the primary (average $\delta$m=4.4 mag) than catalogue secondaries (average $\delta$m=3.3 mag). This highlights the effectiveness of adaptive optics observations over large-scale surveys for detecting new, close and faint secondary sources. \\

\subsection{Eliminating chance projections}
\label{sec:elimchance}
It is important to discern whether the detected secondary sources are physical companions or simply visual binaries due to chance projection. The most conclusive way to settle this is by measuring relative motions with multi-epoch observations. Alternatively with multi-wavelength data, colour-colour and colour-magnitude diagrams can be constructed which may indicate whether secondary sources are located at the same distance as the primaries. \\
As only single-epoch and single wavelength data was available, we employed statistical methods to determine which objects are likely to be physical companions. Densely crowded fields are more likely to give rise to spurious companions. For their sample, \citet{oudmaijer10} argue that the probability of chance projection depends on the surface density of sources  from both the fore- and background (we will further refer to these as ``background stars''). \citet{correia06} add distance from the primary source as a factor in determining the probability of an object being a chance projection. Assuming the distribution of unrelated sources is random and uniform over the observed field, the probability of chance alignment is given by an exponential decay with area from the primary:
\begin{equation}
P=1-e^{-\pi  d^{2} \rho},
\end{equation}
where $\rho$ is the background source density, in arcsec$^{-2}$, and $d$ the separation between the primary and potential companion in arcsec. We determined the background density of sources brighter than the limiting magnitude using the 2MASS point-source catalogue with a square aperture of 1 arcmin centred on the main target. These probabilities of chance projections are presented for all objects in Table \ref{tab:naco:elimcomps}. The average probability of chance projection is 30\%. Most of the sources with low probability are close to their primaries (19/22 sources with P$<$20\% have separations less than 3"), due to the design of the formula. Also, the majority of likely real companions were not found by previous surveys, with 19/22 sources with P$<$20\% being new detections. 
\begin{table}
\footnotesize
\centering
\caption[Probability of chance projections of all detected companions]{Separations from primary, background source densities (BSC) and probabilities of chance projections of all detected companions.}
\label{tab:naco:elimcomps}
\begin{tabular}{llrr}
\hline
Companion name    & Separation       & BSC            & P$_{chance}$        \\
                   &(")   &  (arcm$^{-2}$) &  (\%) \\
                      \hline
G221.9605-01.9926B & 0.6   & 13                      & 0.4                 \\
G221.9605-01.9926C & 0.7   & 13                      & 0.6                 \\
G221.9605-01.9926D & 1.1   & 13                      & 1.4                 \\
G221.9605-01.9926E & 1.2   & 13                      & 1.6                 \\
G232.0766-02.2767B & 3.1   & 12                      & 9.8                 \\
G232.0766-02.2767C &  4.3 & 12                      & 28                \\
G232.0766-02.2767D & 5.6   & 12                      & 18                \\
G232.6207+00.9959B & 6.4   & 37                      & 74                \\
G232.6207+00.9959C & 6.7   & 37                      & 76                \\
G268.3957-00.4842B & 1.9   & 10                      & 3.2                 \\
G268.3957-00.4842C & 4.8   & 10                      & 18                \\
G268.3957-00.4842D & 8.7   & 10                      & 48                \\
G268.3957-00.4842E & 8.9   & 10                      & 50                \\
G282.2988-00.7769B & 1.6   & 14                      & 3.0                 \\
G282.2988-00.7769C & 2.7   & 14                      & 8.6                 \\
G287.3716+00.6444B & 1.3   & 16                      & 2.2                 \\
G287.3716+00.6444C & 1.5   & 16                      & 3.0                 \\
G287.3716+00.6444D & 1.8   & 16                      & 4.5                 \\
G287.3716+00.6444E & 1.9    & 16                      & 4.9                 \\
G290.3745+01.6615B & 0.7   & 24                      & 1.0                 \\
G290.3745+01.6615C & 1.9   & 24                      & 7.1                 \\
G290.3745+01.6615D & 4.1   & 24                      & 29            \\
G293.5607-00.6703B & 5.7   & 8                       & 20                \\
G293.5607-00.6703C & 4.6   & 8                       & 14                \\
G300.3412-00.2190B & 7.4   & 28                      & 74                \\
G301.8147+00.7808A\_B & 2.9 & 26                      & 17                \\
G301.8147+00.7808A\_C & 3.9 & 26                      & 29                \\
G305.3676+00.2095B & 0.9   & 34                      & 2.3                 \\
G310.0135+00.3892B & 2.6   & 31                      & 16                \\
G320.1542+00.7976B & 2.9   & 60                      & 37                \\
G326.4755+00.6947B & 2.2   & 23                      & 9.2                 \\
G327.9455-00.1149B & 7.4   & 17                      & 55                \\
G331.3576+01.0626B & 3.8   & 54                      & 49                \\
G331.3576+01.0626C & 7.8   & 54                      & 94                \\
G331.3576+01.0626D & 11.3   & 54                      & 100                \\
G331.3576+01.0626E & 11.4   & 54                      & 100                \\
G332.9868-00.4871B & 9.1   & 7                       & 40                \\
G334.7302+00.0052B & 5.8   & 50                      & 77                \\
G336.4917-01.4741B\_B & 4.9 & 34                      & 51                \\
G339.6816-01.2058B & 8.6   & 24                      & 79  
\\
\hline             
\end{tabular}
\end{table}
\noindent
Based on this procedure, and using typical values of background source density and separation, it is expected that no more than 20\% of the detected companions at small separations ($<$3") are chance projections (corresponding to 4.4 out of 22 secondaries). As a test, one can look for the number of sources detected at a random location in the image. For example, if the target of the image was not in the upper right quadrant 2, but at the center of the bottom left quadrant 4, how many objects would be found by chance alignment in that quadrant? In order to test this, we calculated the separation of all detected sources (including the target MYSOs) from the centre of the bottom left quadrant, at the opposite side of the image. \\
These can then be plotted onto a histogram to determine whether the prediction that 20\% of the companions are chance projections is accurate. Three objects are indeed found to be located by chance in the opposite quadrant, close to the estimate for spurious binaries, 4.4. This test was repeated for the other quadrants of the image, as the bottom left was affected by the technical fault mentioned in Section \ref{sec:naco:obs}. Similar numbers of chance companions are found by focusing on the other image quadrants. \\

\subsection{Physical companions}
All of this evidence points to the fact that the most likely physical companions are located within the same quadrant of the image as the target (at separations of $<$3") and with a low probability of chance projection (P$_{spurious}<$20\%). Applying these selection criteria results in a master sample of 18 physical companions. Of these, only one was previously detected by 2MASS and none were detected by VVV, again highlighting the value of targeted AO observations over all-sky surveys for binarity. The list of physical companions is presented in Table \ref{tab:naco:physcomps}. \\
We calculated the companion and multiplicity fractions (CF and MF). These are given by the following formulae: MF=$\frac{N_{m}}{N_{t}}$ and CF=$\frac{B+2T+3Q+...}{S+B+T+Q+...}$, where N$_{m}$ is the number of multiple systems, N$_{t}$ is the total number of observed systems, S is the number of single systems, B is the number of binary systems, T is the number of triple systems and Q of quadruple systems. As there are 32 systems, with 10 multiples, of which 6 binaries, two triples, one quadruple and one quintuple, the resulting values for the fractions are MF=31$\pm$8\% and CF=53$\pm$9\%. \\

\begin{table*}
  
\caption[Properties of candidate companions within 3" and with a probability of chance alignment of$<$20\%]{Properties of candidate companions within 3" and with a probability of chance alignment of$<$20\%; Masses and mass ratios with the foreground (fg) and circumstellar (circ) extinction estimates are provided. Considering all the sources of error, mass estimates carry uncertainties of order 30\%, while those on the mass ratios are also estimated to be of order 30\%  (see text for details). P$_{spur}$ 2MASS is the probability of chance projection based on 2MASS source counts.}  \label{tab:naco:physcomps}
\begin{tabular}{llrrrrrrll}
\hline
Name                  & Sep         & Phys.         & P$_{spur}$ & A$_{V\_fg}$ & A$_{V\_circ}$ & M\_fg         & M\_circ      &  q            &  q             \\
                      & (")         & sep(au)       & 2MASS      & (mag)       & (mag)         &  (M$_{\odot}$) &  (M$_{\odot}$) &  fg            &  circ          \\
\hline
G221.9605-01.9926B    & 0.6$\pm$0.3 & 1920$\pm$960  & 0.4        & 1.5         & 38.8          &  9.6   &  52.3 &  1.00 &  5.45 \\
G221.9605-01.9926C    & 0.7$\pm$0.3 & 2240$\pm$960  & 0.6        & 1.5         & 38.8          &  10.4    & 56.8 & 1.10 & 5.92 \\
G221.9605-01.9926D    & 1.1$\pm$0.3 & 3520$\pm$960  & 1.4        & 1.5         & 38.8          &  12.3  & 67.0   & 1.28 & 6.98 \\
G221.9605-01.9926E    & 1.2$\pm$0.3 & 3840$\pm$960  & 1.6        & 1.5         & 38.8          &  7.8   & 42.5 & 0.81 & 4.43 \\
G232.0766-02.2767B    & 3.1$\pm$0.2 & 9390$\pm$600  & 9.8        & 1.8         & 26.7          &  10.8  & 33.6 & 1.16 & 3.62 \\
G268.3957-00.4842B    & 1.9$\pm$0.1 & 1344$\pm$70   & 3.2        & 0.2         & 59.4          &  1.6    & 23.5  & 0.20 & 3.03 \\
G282.2988-00.7769B    & 1.6$\pm$0.1 & 5809$\pm$370  & 3.1        & 2.8         & 23.3          &  3.1    & 7.8   & 0.36 & 0.91 \\
G282.2988-00.7769C    & 2.7$\pm$0.1 & 10027$\pm$370 & 8.6        & 2.8         & 23.3          &  2.5    & 6.3   & 0.29 & 0.74 \\
G287.3716+00.6444B    & 1.3$\pm$0.1 & 5715$\pm$450  & 2.2        & 1.5         & 25            &  7.0     & 20.4  & 0.47 & 1.36 \\
G287.3716+00.6444C    & 1.5$\pm$0.1 & 6660$\pm$450  & 3.1        & 1.5         & 25            &  4.2   & 12.4  & 0.28 & 0.83 \\
G287.3716+00.6444D    & 1.8$\pm$0.1 & 8190$\pm$450  & 4.5        & 1.5         & 25            &  2.6    & 7.5   & 0.17 & 0.50 \\
G287.3716+00.6444E    & 1.9$\pm$0.1 & 8550$\pm$450  & 4.9        & 1.5         & 25            &  5.9   & 17.2    & 0.40 & 1.15 \\
G290.3745+01.6615B    & 0.7$\pm$0.1 & 2030$\pm$290  & 1.2        & 2.2         & 21.5          &  8.0     & 19.3  & 0.57 & 1.38 \\
G290.3745+01.6615C    & 1.9$\pm$0.1 & 5452$\pm$290  & 7.1        & 2.2         & 21.5          &  5.7   & 13.8    & 0.41 & 0.99 \\
G301.8147+00.7808A\_B & 2.9$\pm$0.1 & 12672$\pm$440 & 17.2       & 2.6         & 42.2          & 3.7   & 22.5  & 0.23 & 1.40 \\
G305.3676+00.2095B    & 0.9$\pm$0.2 & 3520$\pm$800  & 2.3        & 2.7         & 45.7          & 5.0     & 39.8   & 0.35 & 2.78 \\
G310.0135+00.3892B    & 2.6$\pm$0.2 & 8192$\pm$640  & 16.2       & 1.8         & 45.7          & 7.6   & 60.6 & 0.32 & 2.57 \\
G326.4755+00.6947B    & 2.2$\pm$0.2 & 3942$\pm$360  & 9.2        & 1.9         & 52.6          & 1.9    & 19.1  & 0.22 & 2.21 \\
\hline
\end{tabular}
\end{table*}

\section{Discussion}
\label{sec:naco:discussion}
\subsection{Multiplicity and companion fractions}

We investigated the variation of the multiplicity and companion fractions across distance, luminosity or extinction. The multiplicity fraction does not differ significantly over distance and luminosity ranges or degree of embeddedness: there are 8 MYSO primaries within $<$2.5 kpc, and 24 systems further away than 2.5 kpc. The multiplicity fractions are 25$\pm$15\% and 33$\pm$10\% for the near and far MYSOs, respectively, so agreeing with each other within the errors. There are 16 MYSOs with a bolometric luminosity under 10,000 L$_{\odot}$, and another 16 brighter than 10,000 L$_{\odot}$. For both of these ranges the multiplicity fraction is 31\%. Finally, there are 16 MYSOs with an A$_{V}$ (estimated from H-K photometry following the method of \citealt{Cooper2013}) of less than 40 magnitudes, and 16 with extinction over 40 magnitudes. The multiplicity fraction for both of these ranges is also 31\%.\\
The overall multiplicity and companion fractions are lower than the values quoted in the literature for massive stars and some low mass young stars. T Tauri stars are reported to have CF=64-79\% depending on their mass range (\citealt{kraus11}) for the separation range 3-5000 au, whereas Class I embedded protostars have CF=37\% according to \citet{connelley08b}. O and B Main Sequence stars have been found to have CF=130 and 100\% and MF=70 and 52\% respectively (\citealt{sana12}) for separations 2-200 au. \\
However, the separation and mass ratio ranges probed by the NaCo data is significantly different to that of other surveys. The separation range of the data presented here is 600-10000 au. \\
The survey of \citet{kraus11} searched for binary companions to low-mass YSOs within 3-5000 au with $\Delta K\approx$6 mag, corresponding to q$_{min}$=0.08.  For a separation and mass ratio range of 600-5000 au and 0.13 respectively (comparable to that of our survey, see also next section), the T Tauri multiplicity fraction is 11.4$\pm$3.2\% and the companion fraction 12.2$\pm$3.3\%. In the NaCo survey presented here, 4 binaries and one quadruple system are found between 600-5000 au, which correspond to MF=16$\pm$7\% and CF=22$\pm$8\%. Both the MF and CF of MYSOs determined from NaCo are higher than those of \citet{kraus11} within the same range, continuing the field star trend of multiplicity increasing with mass (for the same evolutionary stage). \\
For class I embedded protostars, taking the data from the \citet{connelley08b} survey over 600-5000 au and q$>$0.13 (a similar range to that probed by our NaCo data), MF=17\%$\pm$3.7\% and CF=18.5\%$\pm$4.1\%. The multiplicity and companion fractions of MYSOs are similar to results for class I embedded low-mass YSOs. This may point to primordial binary fractions being the same across the mass range, with dynamical interaction over the formation process and subsequent evolution resulting in the observed differences in multiplicity across the mass range for MS stars. However, it is worth pointing out that these surveys probe a larger sample than this NaCo pilot survey. Also, the Taurus cluster probed in this survey is known to have an unusually large multiplicity fraction compared to other low-mass star forming regions, so it may be that the MYSO multiplicity is larger than that of low-mass class I YSOs. \\
\citet{turner08} surveyed O stars in the $I$ band searching for wide companions. Applying the constraints of the NaCo data survey (separations 400-46,000 au and q$>$0.12) to their findings results in MF=17\%$\pm$3.8\% and CF=23\%$\pm$4.3\%. These values are lower than the multiplicity and companion fractions of MYSOs from the NaCo data (MF=31\%, CF=53\%). This lends support to the idea that MS multiplicity fractions are lower than primordial fractions due to dynamical evolution. \\
In conclusion, and with the caveat that it is not trivial to directly compare all results, the multiplicity fraction of MYSOs is larger than that of lower-mass Class II/III YSOs and than that of MS O stars over a similar range in separations and mass range. The multiplicity of embedded Class I  YSOs (of a lower mass and at a potentially earlier evolutionary stage) is similar to the multiplicity fraction of MYSOs. This suggests that multiplicity fractions increase with mass for objects of the same age and decrease with evolutionary stage for objects of the same mass due to dynamical interactions.

\subsection{Masses and mass ratios}

\begin{figure*}
\centering
\includegraphics[scale=0.3]{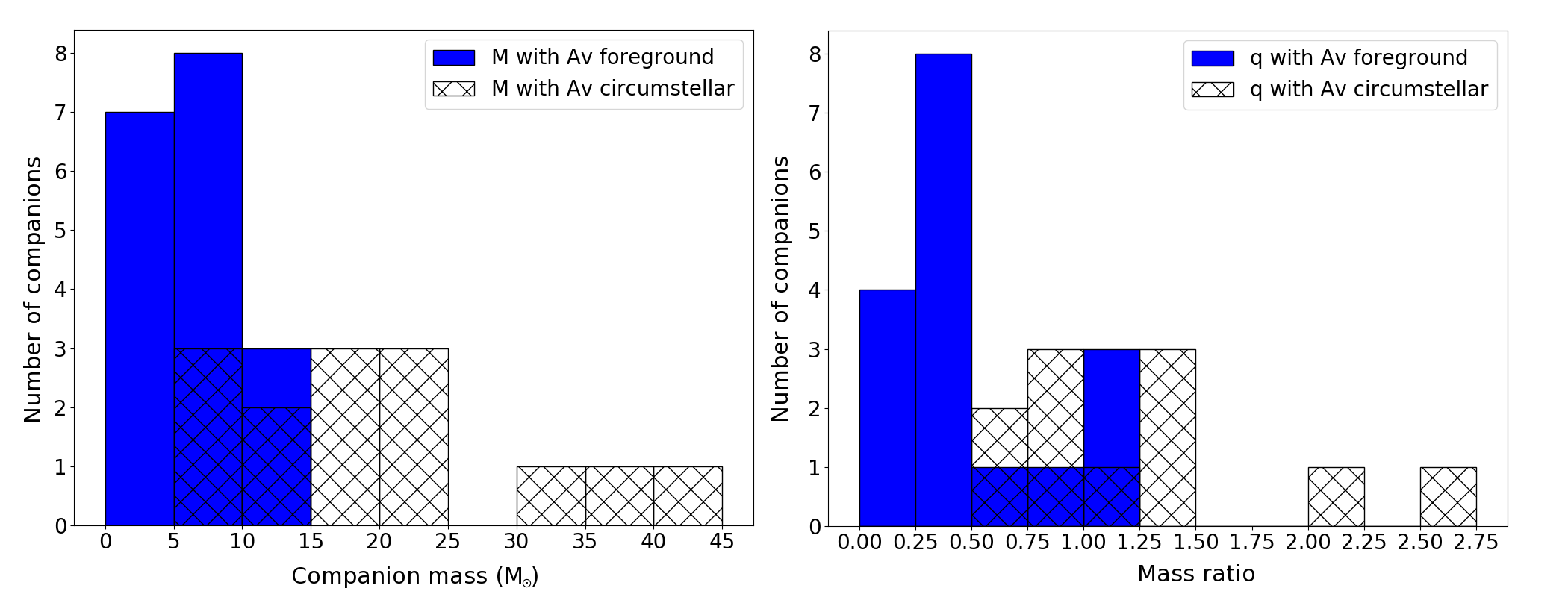} 
\caption{\small Histograms of companion masses (left) and mass ratios to the primary (right). The two distributions correspond to the two different limits of estimating extinction to the secondary. For the blue distribution, extinction was assumed to be just the foreground component, whereas for the hatched distribution we assumed the A$_{V}$ was the same as to the primary, the circumstellar extinction.}
\label{fig:naco:massratios}
\end{figure*}

In principle, the ratio of the $K$ band magnitudes can be used as a proxy for the stellar mass of companions in a multiple system, assuming they are both on the MS, as for example done by \citet{oudmaijer10} in the case of Be stars. However, this applies to field stars unaffected by differential dust excess. There are a number of caveats that must be considered when calculating the mass ratios. The NaCo data presented here covers separations of order 10$^{3}$ au, so it is very likely that the differential excess does play a role. Companions at an earlier evolutionary state than the primary target may be more embedded. Additionally, strong accretion produces strong excess emission. Unfortunately, no multi-wavelength data is available to help quantify the embeddedness of the companions. \\
In spite of all these caveats, an analysis of the limits of mass ratios can provide insights on the composition of multiple young systems. \\
We estimate primary masses using the RMS bolometric luminosities and mass-luminosity relations for MYSOs presented by \citet{Davies2011}, the mass ratio is then computed using the mass of the companion that is determined independently of the primary, as detailed below.  \\
In order to calculate the mass of the companions, their measured $K$ band magnitude is first converted to absolute by adopting the RMS distance to the primary as the distance to the system. Next, a correction for dust extinction is applied. For tight binaries, the extinction to the primary (which we will refer to as the total extinction for clarity from here) can be used, as the whole system may be embedded in the same dust cloud. In the case of wide binaries, which is more likely to be the case here,  the companion may not be shrouded by the same amount of dust as the primary, and as such the total extinction is likely to be inaccurate. The lower limit to the extinction is the foreground extinction at the given distance for the Galactic line of sight the system is located in.\\
\citet{neckel80} provide maps of foreground extinction for most of the Galaxy, and we adopt these as lower limits for the A$_{V}$. The total extinction of the primary, consisting of the interstellar and circumstellar extinction,  can then be used as an upper limit. The total A$_{V}$ is determined by comparing 2MASS $H-K$ photometry to the expected colours of a MS B0 star, as shown by \citet{Cooper2013}. The corrected absolute $K$ band magnitudes can then be used to determine the masses of the companions, by using the equation of \citet{oudmaijer10} (for MS stars):
\begin{equation}
\label{form:massrat}
log(M / M_{\odot}) = -0.18 K_{abs} + 0.64
\end{equation}
\noindent \\ As such, two limits to the mass of the companion can be determined: a lower limit by correcting the companion magnitudes with the foreground extinction and an upper limit by using the total extinction. The mass of the companion will likely be somewhere between these two values. A histogram of the resulting companion masses
(determined from the absolute $K$ band magnitudes) and mass ratios, for both of the methods of estimating extinction is shown in Figure \ref{fig:naco:massratios}. The average mass of the companions is 6 M$_{\odot}$ (corresponding to a B2.5V star) when using foreground extinction and 29 M$_{\odot}$ with total extinction. The mass ratio
averages are 0.5 for foreground extinction and 2.3 for total extinction. This points out to the difficulty in using the same extinction for the companion as for the primary star, as this results in companions more massive than the MYSOs for most systems. The disagreement is likely caused by the wide separation of the binaries reported here. As explained above, wide secondary components may have lower amounts of extinction than the primaries, and so using the total extinction results in an overcorrection of the secondary magnitude.\\ 
It is worth mentioning some important caveats regarding the simplified determination of the secondary masses, and therefore the mass ratios, using a single $K$-band magnitude measurement. The main assumption behind this mass determination is that the $K$-band brightness, like the total luminosity, remains constant on the pre-main sequence evolutionary tracks, and can therefore act as a proxy for the mass. Given that these tracks concern a temperature evolution, the $K$-band magnitude will be brighter in the earlier phases where the objects are cooler, resulting in a too large assignment of the mass.  An additional complication is that large $K$ band excess emission of the secondary, either due to accretion or dust, would result in a lower fraction of the $K$ band magnitude being due to direct photospheric emission from the companions themselves, so the secondary masses, and the mass ratios could also be lower than determined here. Having said that, excess emission would be accompanied by dust extinction, complicating the matter even more. To summarise, the shape of the relationship between mass and $K$ band magnitude may be different to what Equation \ref{form:massrat}
suggests. An investigation into this relation would require an independent measurement of $K$ band excess, perhaps from spectral energy distribution fitting. This analysis is however, beyond the scope of this current work. Further data will certainly be useful to constrain the masses even more.\\
With regard to the (formal) uncertainties on the mass ratios, these are estimated to be of order 30\%. Although the uncertainty in the distances may be the largest contributor to the uncertainty in the mass of the individual components, they essentially cancel out when taking the mass ratio. This is because both the mass determinations depend to first order in a similar manner on the distance. This leaves - methodology aside - the uncertainty in the bolometric fluxes for the primaries (estimated by \citet{mottram11}  to be of order 10-20\%), the {\it K}-band photometry of the secondaries (of order 0.1-0.2 magnitudes in Table~2) and the assumed extinction to the secondaries as main, formal, contributions to the error budget in the luminosities. Given that $A_K$ is around one tenth of $A_V$, this means that for the foreground extinctions which are of order 1.5-2 magnitudes in $V$, even a factor of 2 error will have a small effect on the final luminosity and mass determination. For the larger extinctions in the case where we assume the circumstellar extinction of the the primary to be also applicable to the companions, errors of order a factor of two will have an effect of order 50\% on the masses, however, as will be discussed later, these masses are likely to be overestimated in the first place. On balance, we will assume an error of 30\% on the mass ratios.\\ 
With these, inevitably many, caveats in mind, we find that a large fraction of companions with high mass ratios ($>$0.5) is found even when using the foreground extinction correction for the magnitudes. It is worth noting that the mass ratios obtained with this method are lower limits, so the real mass ratios are likely larger. The number of high mass companions is thus much larger than one would expect if the companions were randomly drawn from the IMF, as predicted by the binary capture formation scenario. The average primary mass of our MYSOs is 14 M$_{\odot}$, so a mass ratio of 0.5 corresponds to a secondary of 7 M$_{\odot}$. Based on the IMF of \citet{salpeter55}, one would expect that there are $\approx$23 times more stars between 1-7 M$_{\odot}$ (so with q$<$0.5) as between 7-14 M$_{\odot}$. However, we find 10 companions between 1-7 M$_{\odot}$ and 8 between 7-14 M$_{\odot}$, inconsistent with random capture. Let us compare this with the MS results by \citet{moe2017} who find larger mass ratios for close binaries but mass ratios consistent with the IMF at the separations probed here. As we seem to find high mass ratios, this leaves the intriguing possibility that the   separation distribution of massive companions evolves. Perhaps high mass companions are formed at larger distances to migrate closer to the primary later, or by triple decay and dynamical hardening or a combination of both.  \\

\begin{table*}
\centering
\caption[Binary and disc position angle measurements]{Binary and disc position angle measurements. * - disc position angle deduced from outflow or jet PA. ** - outflow seen as extended emission in 2MASS image}
\begin{tabular}{llll}
\hline
Object name           & Binary\_PA ($^{\circ}$)& Disc\_PA ($^{\circ}$) & Reference  \\                  
\hline
G221.9605-01.9926B    & 75         & 30*      & \citet{zhang05}       \\
G232.0766-02.2767B    & 178        & 175*     & \citet{navarete15}    \\
G268.3957-00.4842B    & 170        & 120*     &  2MASS** \\
G282.2988-00.7769B    & 126        & 80*      &  \citet{navarete15}         \\
G290.3745+01.6615B    & 153        & 135*     & \citet{gredel06}      \\
G301.8147+00.7808A\_B & 115        & 65*      & 2MASS**          \\
G310.0135+00.3892B    & 41         & 45       & \citet{kraus10}      \\
G326.4755+00.6947B    & 179        & 125*     & \citet{navarete15}   \\
\hline
\label{tab:naco:PA}
\end{tabular}
\end{table*}

\subsection{Alignment with discs}
\label{sec:naco:discalign}
\begin{figure}
\centering
\includegraphics[scale=0.29]{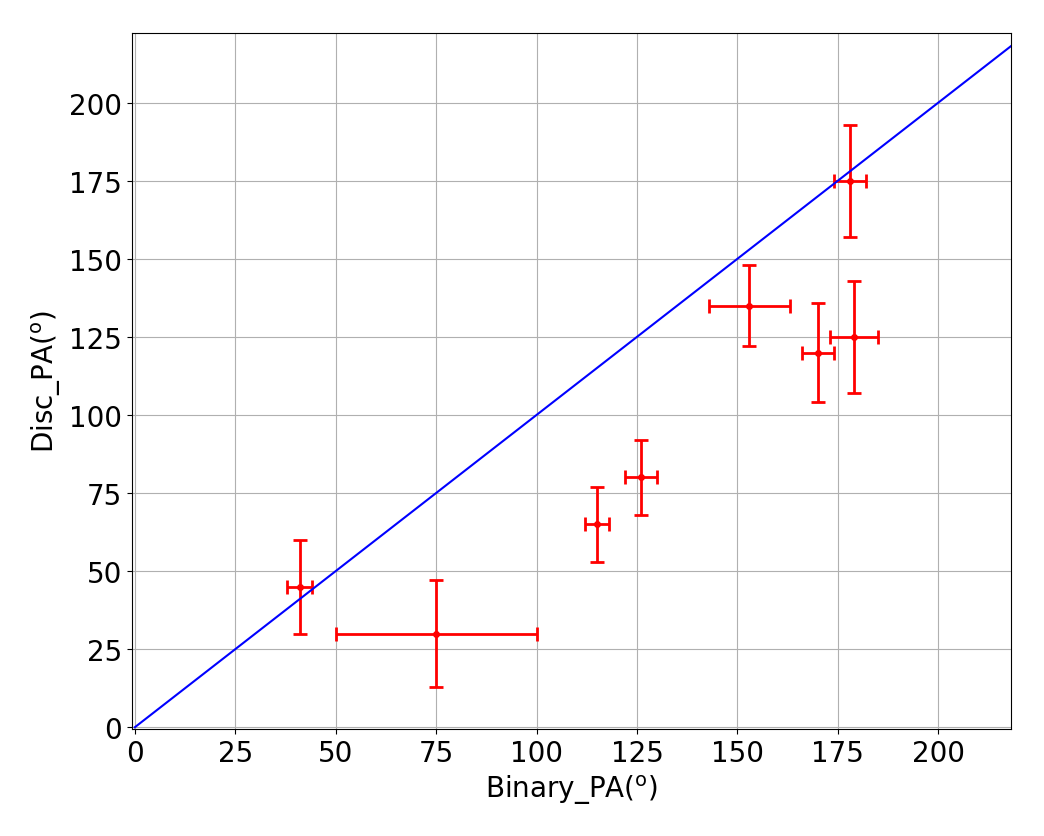} 
\caption{\small Position angle of the disc as a function of the position angle of the binary companion. The blue solid line is the 1-1 correlation, for equal disc and binary orientations.}
\label{fig:PA}
\end{figure}

\begin{figure}
\centering
\includegraphics[scale=0.31]{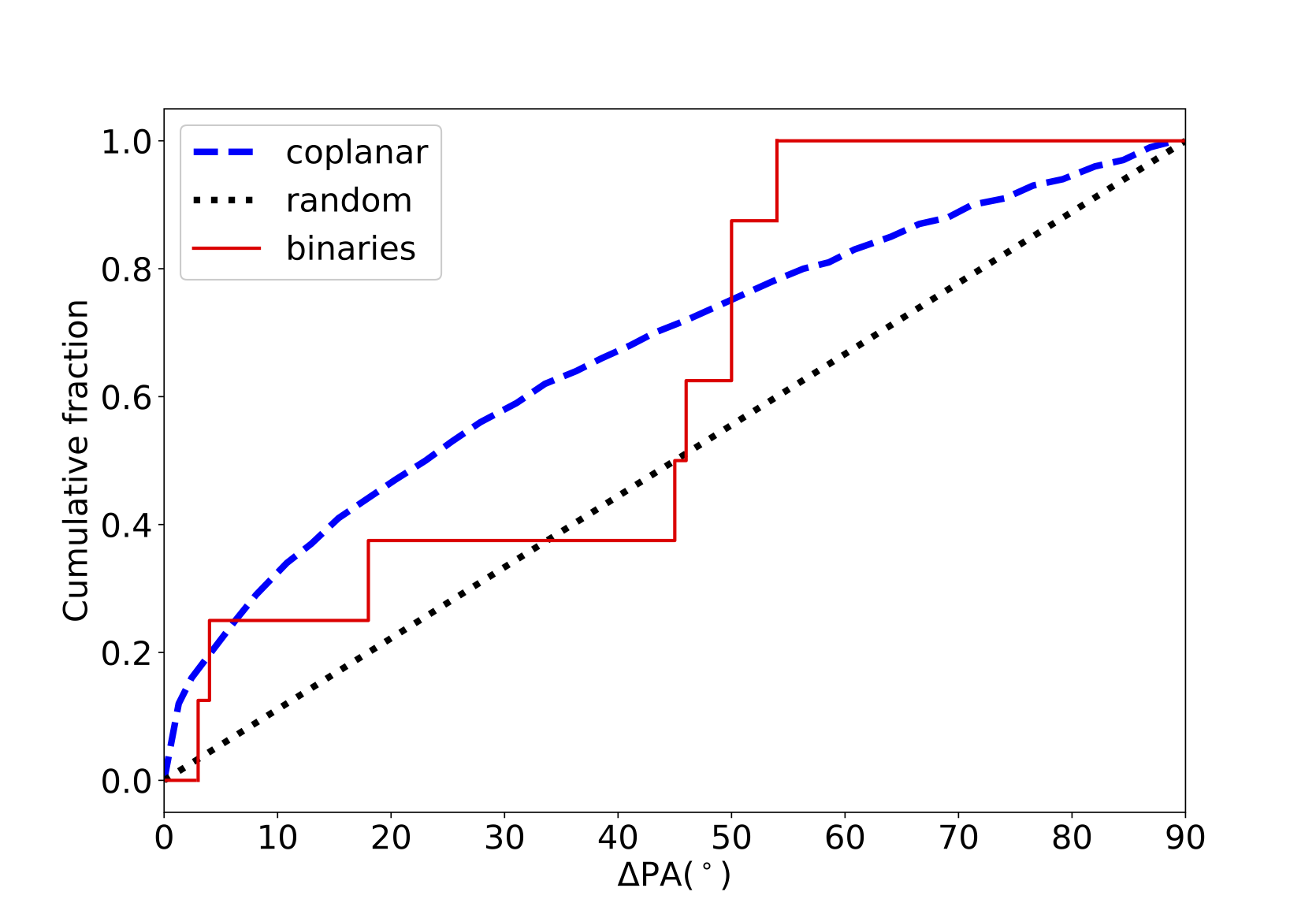}
\caption{\small Comparison of the distribution of the disc-binary PA (red solid line) with the simulated distributions of  \citet{wheelwright11} for coplanar (blue dashed line) and random (black dotted line) orbits}
\label{fig:PAcomp}
\end{figure}

As mentioned in Section \ref{sec:naco:intro}, one can test the different binary formation models by comparing the alignment of the accretion disc with that of the binary orbit. If the secondaries were formed as a result of the fragmentation of the accretion disc, the orbit of the companion should be located within the same plane as the accretion disc. Arguably, the orbit of the companion is unlikely to have strayed significantly from the plane of the accretion disc in the short time from the formation of the companion to the MYSO phase. However, this is complicated by the unknown angle at which we are viewing the binary system. For edge-on systems (or at large viewing angles), an aligned disc-companion configuration will indeed result in the binary orbit and the accretion disc having the same position angle (PA). For angles at lower inclinations with respect to our line of sight, the disc and binary orbit may appear to be at different position angles even if they are within the same plane in reality. \citet{wheelwright11} surveyed the multiplicity of Herbig stars with spectro-astrometry. They used a model to predict the cumulative distribution function of the difference between the disc and binary PA when the orbits are coplanar and when the PAs are distributed randomly, comparing the observed disc-binary orbit PAs to these two different distributions. With this they showed that binary rotation axes and protostellar discs of Herbig stars are consistent with being aligned at a 2.2$\sigma$ level, as predicted by the disc fragmentation binary formation theory. \\
Such an analysis is more complicated for MYSOs, first of all due to their lower relative numbers compared to Herbig stars - at least 100 measurements are required for a 3-$\sigma$ precision, and 20 for 2$\sigma$. In addition, direct disc detections are rare. The other option for inferring the disc PA is modelling of disc tracers or other elements of the circumstellar environment. Finally, as outlined in the Introduction, detections of binaries in MYSOs are also rare. \\
We used the RMS database, 2MASS images and literature to investigate the presence of discs, outflows or jets in our targets and their position angles. If a jet, outflow or extended emission was detected, it was assumed the disc PA would be oriented at 90$^{\circ}$ with respect to the PA of the outflowing gas. This is what most high-resolution observations of disc-outflow massive young systems find (eg. \citealt{gibb07}). The angles are normalised to the (0,180)$^{\circ}$ range. Disc PA measurements are available for 8 of the detected companions. The data is shown in Table \ref{tab:naco:PA}. We note that \citet{ababakr17} also observed a disc in this object through spectropolarimetry, inclined at 146 $\deg$. However, in the interest of consistency with the previous measurement, we use the disc inclination of 80 $\deg$ from \citet{navarete15}.\\
There is a weak correlation between the disc and the binary PA, with a Pearson correlation coefficient of 0.71, corresponding to a probability of false correlation of 0.1\%. The plot is displayed in Figure \ref{fig:PA}.\\
We also compared the measured distribution of the difference between the disk and binary PA to the simulations of \citet{wheelwright11} in Figure \ref{fig:PAcomp}. This data cannot distinguish between the two simulated distributions, with the distribution being located at equal distance from the coplanar and random distributions. The observed data are best fitted by the random distribution at low disc-binary PAs, and by the coplanar distribution at large disc-binary PAs. This is likely due to the smaller size of the NaCo data set compared to the sample of \citet{wheelwright11}. Their sample contained 20 Herbig Ae/Be stars, whereas the NaCo sample only has 8 companions with disc PA measurements available. \\

%\label{sec:PA}

\subsection{Are MYSOs with binaries special?}
\label{sec:PA}

\begin{figure}
\centering
\includegraphics[scale=0.33]{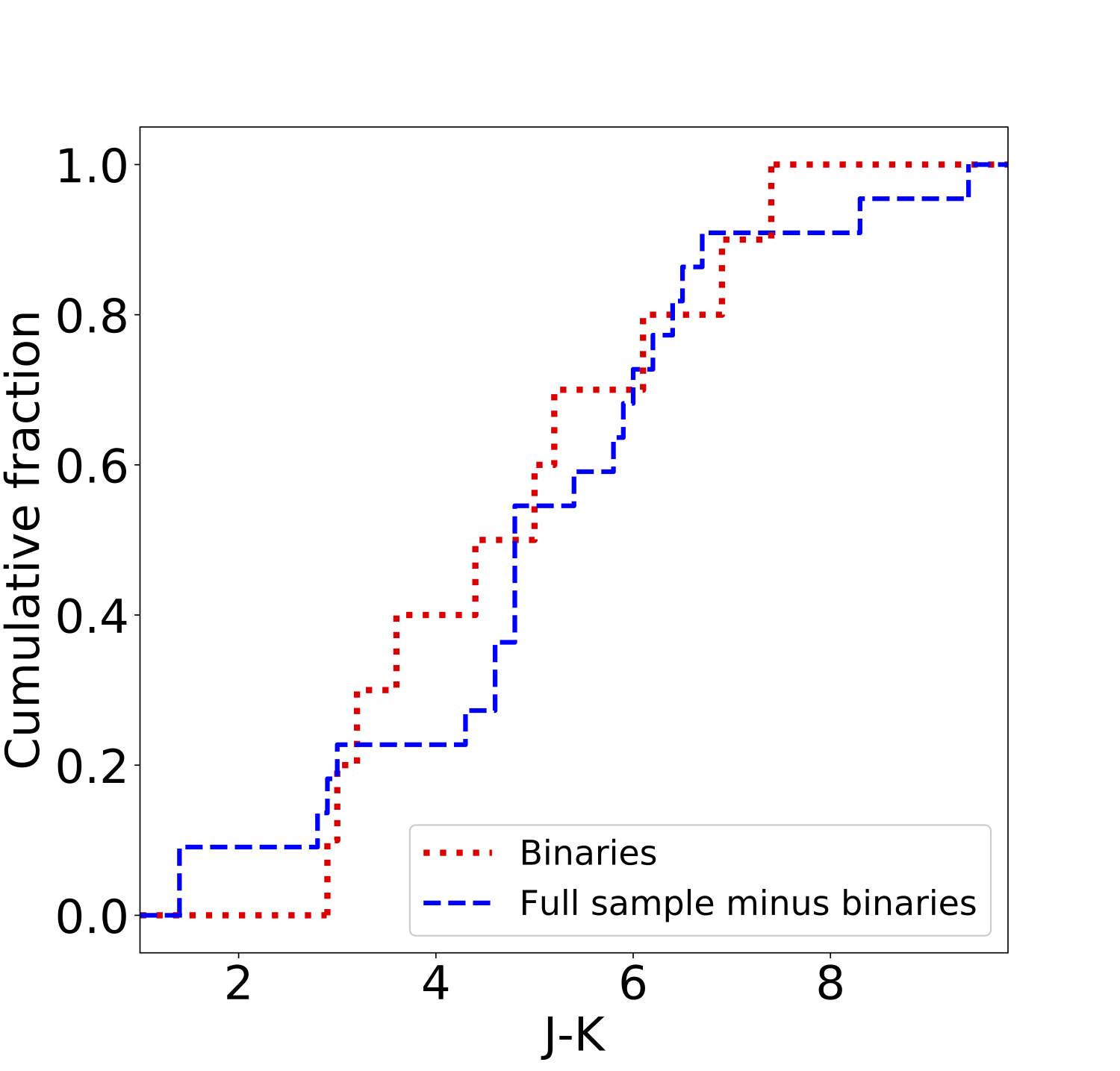} 
\caption{\small Comparison of the distribution of the J-K colour of MYSOs in our sample with binaries (red dotted line) and the whole NaCo sample (blue dashed line).}
\label{fig:JKcomp}
\end{figure}

Following the approach of \citet{ilee13}, we study the sample of MYSOs with binaries in order to determine whether their properties differ from the properties of single MYSOs. \\ 
The average luminosity of binary MYSOs is 16,000 L$_{\odot}$, and the average distance is 3.1 kpc. This is similar to the average luminosities and distances of the whole NaCo sample, 15000 L$_{\odot}$ and 3.3 kpc respectively. The averages for all the RMS MYSOs are 11500 L$_{\odot}$ and 4.5 kpc. The difference in properties between the binaries and the whole RMS sample can be explained by the initial sample selection criteria of L$>$3500 L$_{\odot}$ and d$<$5 kpc. As such, the whole RMS database contains a large number of fainter and more distant MYSOs than this NaCo sample. When restricting the full RMS MYSO sample to L$>$3500 L$_{\odot}$ and d$<$5 kpc, the average bolometric luminosity and
distance of MYSOs are 20500 L$_{\odot}$ and 3.3 kpc respectively. Histograms of the luminosity and distance distributions of the MYSOs with binaries appear visually similar to the same distributions for the NaCo sample without detected binaries. K-S tests were applied, and are consistent with this result. The luminosity distribution yielded a
K-S statistic of 0.32, indicating that the probability of the binary sample to be drawn from the same distribution as the complete sample is 68\%. The K-S test between the binary and the rest of the NaCo sample for the distance distribution resulted in a probability of 78\% that the two distributions are drawn from the same initial sample. Therefore, based on the results from this data set, there is no obvious difference between MYSOs with or without companions in terms of bolometric luminosities or distance. However, this conclusion is limited by the small size of the studied sample and the limitations of the K-S test. \\ 
We also searched for outflows or discs in the objects in the sample. 60\% of MYSOs with companions and 40\% of those with no companions have an outflow or disc. As such, we conclude that there are no differences between the binary and single MYSOs in the NaCo sample in terms of outflow or disc detections.\\
Finally, we checked whether there are any biases of the sample of objects with physical companions in terms of the amount of dust. We compared the embeddedness of MYSOs with binaries to the other objects in the NaCo sample by using the $J-K$ colour as a proxy for the amount of dust. Cumulative distribution functions are shown in Figure \ref{fig:JKcomp}, and K-S tests show that the probability of the two samples being drawn from the same initial distribution are larger than 82\%. The A$_{V}$s of the MYSOs with and without binaries are also very similar according to a K-S distribution test. The statistic of 0.17 corresponds to a probability of 98\% that the two distributions are drawn from the same sample.\\
Based on these observations, we conclude that the MYSOs with and without binaries are not extremely different. 

\subsection{Estimating the total multiplicity and companion}
Taken at face value, our results are that 1-in-3 MYSOs have at least one companion with a separation on the sky of about $10^3$ -- $10^4$\,au, and a CSF in this range of roughly a half. \\
Our observations are obviously limited in a number of ways. The sample size is relatively small, and so the errors on any numbers are large. The observations are not uniform, with varying selection effects depending on the particular conditions of any particular observation.  With the problem of unknown (and presumably varying) extinction to the secondaries the masses and any mass ratios are extremely difficult to determine.  Despite these limitations we think that it is quite reasonable to draw the following three conclusions from the data.\\
{\bf The fraction of MYSOs in multiple systems is close to 100 per cent.} The results from this analysis are that there are about 30 per cent of MYSOs with a companion we can observe.  For a companion to be observable it must be wide (between about $10^3$ and $10^4$\,au), and it must be of similar brightness -- hence mass -- to the primary.  Determining masses to any degree of precision is extremely difficult, but the mass ratios of the systems we observe are probably all $q>0.5$ (ie. a companion of at least half the mass of the primary).\\
So our rough limits are observing only companions with separations $10^3$ -- $10^4$\,au with $q>0.5$ and we find a multiplicity fraction of about 0.3 ($1 \sigma$ limits between about 0.2 and 0.4). Extrapolating over the whole range of separations and mass ratios should increase the multiplicity fraction by factors of several.\\  
Observations of MS A-stars between 30 and 45000\,au by \citet{derosa14} find that the separation distribution peaks at $\sim 400$\,au.  These authors also find that the mass ratio is biased to lower mass companions especially at large separations (which matches the B star observations of \citealt{kouwenhoven05}).  This suggests we are missing well over half of all companions as they are within our resolution limit (ie. are $< 10^3$\,au), and even in the range we can see we are missing well over half of companions as they are too faint to observe.  If we take the {\em very} conservative limits that we are complete between $10^3$ -- $10^4$\,au with $q>0.5$, and only missing half of stars because of the limited separation range, and half because of the mass ratio limit then the underlying multiplicity fraction is at least $0.3 \times 2 \times 2 = 1.2$, more reasonable extrapolations increase this to $> 2$ which leads us to our next conclusion.\\
{\bf Many (maybe all) MYSOs are higher-order multiple systems.}  We are finding a significant number of companions $>10^3$\,au, if the peak of the MS separation distribution is indeed at around 400\,au this suggests we are tending to observe the outer component of higher-order multiple systems.  That in the limited range of separations and masses we are sensitive to we observe 2 triple systems, a quadruple and a quintuple is strong support for this idea, as is the apparently high higher-order multiplicity fraction in MS A-stars (nearly 50 per cent according to \citealt{derosa14}).  We note that some/many of these young higher-order systems will be unstable and could quite plausibly decay into a population that looks very similar to the MS A-star distribution.\\
{\bf Many MYSOs are the most massive stars locally.}  Companions at distances of $10^3$ -- $10^4$\,au from the primary are quite `soft' in that they are relatively easy to destroy.  An encounter will destroy a $10^3$ -- $10^4$\,au system if it carries the same, or more, kinetic energy than the binding energy of the system.  For fairly typical $15 + 10 M_\odot$ system at $10^3$\,au the encounter velocity required to destroy the system depends on the mass of the perturber stars (in solar masses) as $\frac{16.43}{m^{-1/2}}$km s$^{-1}$ .  A $1 M_\odot$ perturber would have to travel at a relatively fast 16.43 km s$^{-1}$, but a $10 M_\odot$ perturber would destroy the system if it encountered it at 5.27 km s$^{-1}$ which is a perfectly reasonable relative velocity for such a star to have in a star forming region. \\
That at least a third of MYSOs have companions we can observe at $10^3$ -- $10^4$\,au suggests that {\em at least} one third of MYSOs have never encountered a similar- or higher-mass star and so must have {\em always} been the most massive objects in their locality (see \citealt{griffiths18} for more details and a very similar argument
about O-stars in Cyg OB2).

\section{Conclusions}
\label{sec:naco:concl}
We have presented AO-assisted $K$ band observations of 32 MYSOs searching for new binary companions. The observations are complete to a contrast of $\Delta K$=5 mag at 1-3" and  $\Delta K$=3 mag at 0.3". This corresponds to a physical separation range of 600 - 10,000 au, within the predictions of models and observations for multiplicity of MYSOs. Statistical methods based on background source density and separation are employed to determine the likelihood of the companions being physical rather than visual binaries. The main findings are as follows:
\begin{itemize}
\item The multiplicity fraction is 31$\pm$8\% and the companion fraction 53$\pm$9\%. These fractions are lower for MYSOs than the overall fractions for T Tauri or Main Sequence O stars. However, for similar separation and mass ratio ranges, the multiplicity fraction of MYSOs is larger than that of T Tauri or O stars. This lends support to theories suggesting multiplicity increases with mass and decreases with evolutionary stage.
\item Lower limits to mass ratios are generally$>$0.5, which is larger than what is expected from randomly sampling the IMF, as the binary capture formation predicts.
\item Due to the low number of sources with disc orientation measurements, this data set cannot differentiate between binary orbits being coplanar to discs or at random orientations.
\item MYSOs with binaries do not show any different characteristics to the average MYSO in terms of luminosity, distance, outflow or disc presence. 
\item From basic considerations, we conclude it is likely that the total multiplicity fraction of MYSOs is close to 100\%, while most of those will reside in high order multiple systems. 
\end{itemize}

These data constitute the first attempts at a systematic study of multiplicity of MYSOs. Multi-wavelength observations will be of great use to fully determine the properties of the companions, while higher spatial resolution data should close the parameter space to smaller separations.   

\section*{Acknowledgements}

We thank Stuart Lumsden and Evgenia Koumpia for their help when preparing the paper and many stimulating discussions.  RP gratefully acknowledges the studentship funded by
the Science and Technologies Facilities Council of the United Kingdom.  Based on observations collected at the European Organisation for Astronomical Research in the Southern Hemisphere under ESO programme 096.C-0623(A). We also make use of the SIMBAD data base, operated at CDS, Strasbourg, France. This paper made use of information from the Red
MSX Source survey database at http://rms.leeds.ac.uk/ which was constructed with support from the Science and Technology Facilities Council of the UK. This publication makes use of data products from the Two Micron All Sky Survey, which is a joint project of the University of Massachusetts and the Infrared Processing and Analysis Center/California Institute of Technology, funded by the National Aeronautics and Space Administration and the National Science Foundation. Based on data products from observations made with ESO Telescopes at the La Silla Paranal Observatory under programme ID 179.B-2002, taken as part of the VVV survey. We also make use of the SIMBAD data base, operated at CDS,Strasbourg, France. PyRAF is a product of the Space Telescope Science Institute, which is operated by AURA for NASA. This research made use of Astropy, a community-developed core Python package for Astronomy (Astropy Collaboration, 2018).

%%%%%%%%%%%%%%%%%%%%%%%%%%%%%%%%%%%%%%%%%%%%%%%%%%

%%%%%%%%%%%%%%%%%%%% REFERENCES %%%%%%%%%%%%%%%%%%

% The best way to enter references is to use BibTeX:

\bibliographystyle{mnras}
\bibliography{citations} % if your bibtex file is called example.bib

%%%%%%%%%%%%%%%%%%%%%%%%%%%%%%%%%%%%%%%%%%%%%%%%%%

%%%%%%%%%%%%%%%%% APPENDICES %%%%%%%%%%%%%%%%%%%%%

%%%%%%%%%%%%%%%%%%%%%%%%%%%%%%%%%%%%%%%%%%%%%%%%%%

% Don't change these lines
\bsp	% typesetting comment
\label{lastpage}
\end{document}